# Blue light absorption enhancement based on vertically channelling modes in nano-holes arrays

*Guillaume Gomard,[t,§,†] Romain Peretti,[t,§] Ségolène Callard,[t,§] Xianqin Meng,[t,§] Rémy Artinyan,[t,§] Thierry Deschamps,[t,§] Pere Roca i Cabarrocas,[‡] Emmanuel Drouard,[t,§] and Christian Seassal[t,§,]**

[t] Institut des Nanotechnologies de Lyon (INL), Université de Lyon, UMR 5270, CNRS-INSA-ECL-UCBL, France

[§] Ecole Centrale de Lyon, 36 Avenue Guy de Collongue, 69134 Ecully Cedex, France

[‡] Laboratoire de Physique des Interfaces et des Couches Minces LPICM-UMR 7647, CNRS, Ecole Polytechnique, Palaiseau 91128, France

ABSTRACT: We investigate the specific optical regime occurring at short wavelengths, in the high absorption regime, in silicon thin-films patterned by periodically arranged nano-holes. Near-field scanning optical microscopy indicates that the incoming light is coupled to vertically channelling modes. Optical modelling and simulations show that the light, travelling inside the low-index regions, is absorbed at the direct vicinity of the nano-holes sidewalls. This channelling regime should be taken into account for light management in optoelectronic devices.

[*] Corresponding author E-mail: Christian.Seassal@ec-lyon.fr

[†] Present Addresses: Light Technology Institute (LTI), Karlsruhe Institute of Technology (KIT), 76131 Karlsruhe,

Trapping and absorbing the incident light in a semiconductor layer with a thickness around 100 nm is essential for devices like thin-film solar cells. Nanophotonics provides various promising approaches to reach this goal. First, a drastic reflection decrease can be achieved with wavelength-scale surface patterns or sub-wavelength biomimetic structures [1-5]. Second, light scattering or coupling into guided modes is an efficient way to trap the light in thin layers, using dielectric [6-10] or metallic [11-13], ordered or disordered [14] patterns. In the first category, although the properties of the patterns are generally simulated by the Rigorous Coupled-Wave Analysis (RCWA) or the Finite-Difference Time-Domain (FDTD) methods, the behaviour of structural anti-reflectors is too frequently discussed or even explained (see e.g [1-3, 15]) using the Effective Medium Theory (EMT) proposed by Maxwell-Garnet or Bruggeman [5, 16, 17]. Such an averaged refractive index is of course not anymore appropriate in the present case, where the dimensions involved are close to, or even higher than the wavelength [18]. In the second category, the impact of nanophotonic structures on light trapping is generally stressed for the long wavelength range, where the absorption increase is clearly attributed to optical resonances [19]. However, their role in the short wavelength range is often unexplained. Still, various studies demonstrated that such nanophotonic structures also lead to an absorption increase in the UV-blue range, notably through a reflection decrease [20-22]. The behaviour of nanopatterned structures is then not described in a satisfactory way in the short wavelength range; this is a limitation for the design of optimized thin-film solar cells.

In this letter, we demonstrate that in nano-patterned thin absorbing layers, the drastic decrease of the reflection and the substantial increase of the transmission and absorption in the short wavelength range are due to the incident light coupling into vertically guided "channelling" modes mainly located in the air holes. This demonstration, and the subsequent investigation of these channelling modes, will be performed using a methodology combining far field optical and Scanning Near Field Optical Microscopy (SNOM) measurements as well as FDTD simulations. We will focus on hydrogenated amorphous silicon (a-Si:H) layers and on an incident light in the short wavelength range ($\lambda$<500 nm), where the absorption coefficient is high (typically around $6.10^5$ cm$^{-1}$) so that the photons lifetime is short. As a result, resonant modes cannot be excited in this spectral range and the reflection losses are dominating over the transmitted ones.

In order to fabricate the investigated samples, 195 nm thick a-Si:H layers were first deposited onto a borosilicate wafer by PECVD. The optical properties of those unpatterned layers are such that they do not allow light transmission for wavelengths below 470 nm (see Figure 1). Then, samples have been patterned over centimetre square areas, with a 2D square array of nano-holes, by combining a holographic lithography step with a reactive ion etching process [9]. The targeted photonic crystal (PhC) period (*L*) and diameter (*D)* values are respectively 500 nm and 322 nm, which are characteristic of those chosen in the literature [8].

The overall reflection (*R*) and transmission (*T*) spectra of those samples, including both the specular and diffuse components, have been measured in the wavelength range of interest by using an integrating sphere with an unpolarized light. These experiments bring uncertainties within 5%. They were performed under a normal incidence for the transmission, and with an 8° angle of incidence for the reflection. The resulting absorption (*A*) is simply derived by applying *A*($\lambda$)=1-*R*($\lambda$)-*T*($\lambda$).

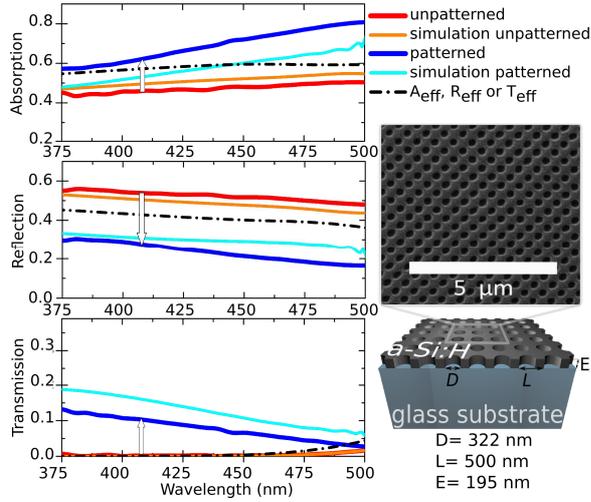

**Figure 1.** Left: Measured (red and dark blue lines) and simulated (orange and light blue lines) spectra of the unpatterned and 2D patterned samples, together with the spectra obtained after averaging the complex permittivity of the patterned layers (dashed lines). Right: Schematic view of the 2D patterned sample with an inset of a scanning electron microscope image of the patterned sample after etching.

As a comparison, the optical properties of those samples were also studied by FDTD simulations. We also calculated analytically the absorption ($A_{eff}$), reflection ($R_{eff}$) and transmission ($T_{eff}$) by considering the patterned layer as flat and homogeneous, and by averaging its complex permittivity by the surface filling factors for air ($ff_{air}=\pi D^2/(4L^2)$) and for a-Si:H ($ff_{a-Si:H}=1-ff_{air}$), similarly to what is used in numerous studies of the literature [1, 23]. This will serve as a reference in order to investigate the specific impact of the patterning, beyond the index matching phenomenon. All these results are displayed in Figure 1, where the data corresponding to the described patterned samples are plotted together with the same but unpatterned sample.

A preliminary remark is that, as no anti-reflection coating is deposited on the test samples, the absorption of the unpatterned reference is lower than 50% over the whole spectral range below 500 nm. Additionally, the measured absorption spectrum is significantly higher than that expected by FDTD simulations. This can be attributed, primarily, to the topographical differences between the simulated and the real fabricated structures (including the roughness of the etched surfaces), and to a lesser extent to the possible error on the optical indices, to the uncertainties of the measurements and to the differences of configuration in reflection mode between the simulations (normal incidence) and the experiment (8° angle of incidence).

One can see on this figure that the reflection is divided by a factor of more than 2 in average due to patterning, giving rise to an absorption increase of about 45%$_{rel}$ when compared to the unpatterned configuration and after integration between 375 nm and 500 nm. Translated in terms of short-circuit current densities, this effect corresponds to an absolute gain of 1.3 mA/cm$^2$ for the same wavelength range. However, the main result concerns the experimental and simulated transmission spectra plotted for the patterned sample which exhibit a radically different trend from the one calculated by the EMT. Indeed, in the latter case, $T_{eff}$ is almost zero up to 450 nm, like in the case of the unpatterned reference. In such situations, the light which is coupled into the layer is totally absorbed in a single pass. On the opposite, the spectra corresponding to patterned samples show that transmission values up to 13% can be reached, with a good agreement between simulation and experiment. This observation can be considered as an extraordinary optical transmission phenomenon, in the sense that it is not predicted by the EMT and does not occur for the unpatterned slab. However, it is important to clarify that the real part of the dielectric function of a-Si:H is positive above 300 nm, so that this patterned layer does not support surface plasmon polaritons.

This transmission increase clearly shows that light propagates deeper into the silicon layer than for the unpatterned samples, which is related to the absorption increase in the patterned ones. To get a better insight into the mechanism of light propagation in such a highly absorbing patterned layer, a local analysis of the optical properties of this nano-scale patterned structure was then performed. To achieve this, transmission mode SNOM experiments were carried out using a stand alone commercial head (NT-MDT SMENA), positioned at the top of an inverted microscope. The sample was illuminated from the transparent borosilicate substrate with a 405 nm laser diode, and the light transmitted through the sample was collected from the top of the patterned layer with a metal-coated fiber tip with an aperture of about 80 nm. This kind of tip is particularly suited when the amount of radiating light cannot be disregarded with respect to the evanescent waves [24]. In this configuration, the probe is sensitive to the optical field at the sample surface, with a sub-wavelength lateral resolution of the order of the tip aperture. The optical map of the field, showing the light modulation over three periods of the PhC, is presented in Figure 2.

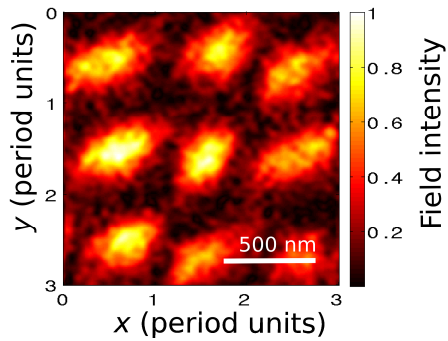

**Figure 2.** Mapping of the measured normalized electrical field intensity on the front side of the patterned sample (*x* and *y*-axis expressed in PhC period units).

It can be concluded here that the intensity is maximal right in the center of the nano-holes, and tends towards a minimal value between two consecutive nano-holes. This peculiar propagation of the light along the nano-apertures in the air region is therefore responsible for the remarkable transmission in the wavelength range of interest. In the following, this optical regime taking place below 500 nm will be referred to as the "channelling regime", indicating that more than 90% of the non-reflected light is propagating vertically in the nano-holes. This behaviour is radically different from that occurring in the case of an unpatterned absorbing layer and in the EMT which assumes a homogenized layer, or for patterned layers in the low absorption range. Another consequence is that the reflection values measured in the channelling regime are lower than the ones calculated with the EMT (see Figure 1) because in the former case, the light in all modes is almost exclusively concentrated in the low index region (Figure 3 left) and hence, this ensures a better impedance matching between the incident light and the absorbing medium.

To corroborate this result, electric field intensity maps were calculated by 3D FDTD simulations. Figure 3 (left) shows the top and cross section views of such structures, as illuminated by a collimated beam at $\lambda$=405 nm.

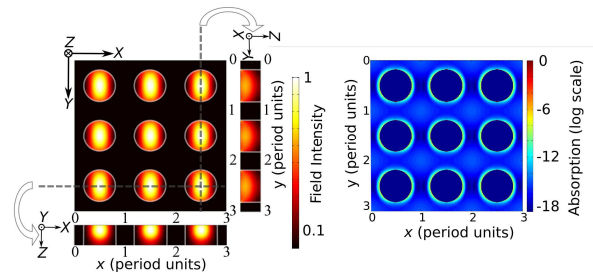

**Figure 3.** Left: Simulated normalized electrical field intensity obtained on the front surface of the patterned layer and for two vertical cuts along the *z* direction. Right: Simulated mapping of the absorption (logarithmic scale). Those mappings are depicted in PhC period units and for a wavelength of 405 nm.

Besides these exact simulation results, a much more simple model can be considered: the case of a single hollow slab waveguide in a highly absorbing high refractive index medium. Such a system can be modelled in a simple way on the basis of the transcendental equations describing a standard slab waveguide, considering a cladding with complex optical constants and a low refractive index core. For example, in the case of TE polarization (with an electrical field along the *y*-axis), and considering the structure depicted in Figure 4, the transcendental equation corresponding to the even mode can be written:

$$k_{x_{a-Si:H}} = -jk_{x_{air}} \tan\left(\frac{d}{2} k_{x_{air}}\right) \quad (1)$$

where *d* is the width of the air slit, and $k_{x_{a-Si:H}}$ and $k_{x_{air}}$ are the *x* components of the wave vector in a-Si:H and in air, respectively (for details see [30]). The electric field intensity map corresponding to this mode is displayed in Figure 4, considering the optical properties of a-Si:H at λ=405 nm. Light is clearly confined in the hollow part of the medium, and is propagating along the *z* direction.

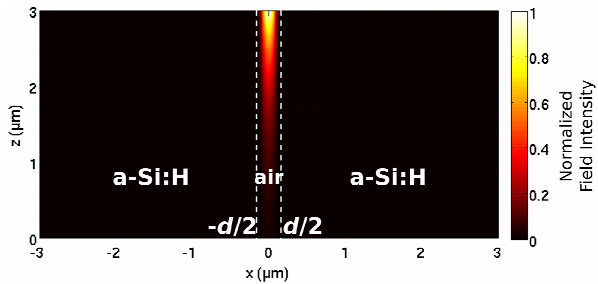

**Figure 4.** Normalized electric field intensity map, calculated at λ=405 nm. The 1D modelled structure corresponds to an air slit inside an a-Si:H layer extending infinitely in the *z* direction.

More generally, this simplified 1D model confirms the existence of TE and TM pseudo-guided modes (the so-called "channelling modes"), corresponding to polarization dependant vertically guided modes in each hole, in the case of the real PhC layer. One should note that the situation is different for the TE and the TM polarizations, since for the TM polarization, the transcendental equations are weighted by the permittivity of the absorbing material, leading to a non-isotropic distribution of the field in the horizontal plane of the membrane. We can also conclude that modes are strongly attenuated in the absorbing medium, which corresponds to the electric field intensity decay in the lateral (*x*) direction, as well as in the propagation direction (i.e. in the *z* direction). This absorption in a very limited volume is also illustrated by the absorption maps displayed in Figure 3 (right). Although the channelling modes are mostly localized in air below 500 nm, the propagation losses due to the absorption of the decay tail in a-Si:H are all the more important that the extinction coefficient of a-Si:H is as high as 1.8 at 405 nm [8], which is enough to enable absorption on several tens or hundreds of nanometers. Thus, assimilating the PhC patterned layer to a series of independent dielectric "waveguides" gives a physical interpretation to the experimental and simulated mappings displayed in Figure 2 and 4, and to the absorption and transmission values measured below 500 nm (Figure 1).

Hence, understanding this channelling regime opens the route to a better exploitation of the blue light so as to improve the external quantum efficiency of PhC patterned photovoltaic solar cells at short wavelengths. In this context, we shall consider a solar cell stack including top and bottom highly doped layers with thicknesses of the order of 10 nm, and an intermediate intrinsic layer with a thickness of the order of 200 nm. Then, it can be concluded that etching the PhC nano-holes beyond (at least) the highly-doped front layer is a well-suited approach to improve the electrical performances of the device, since the channelled photons will be absorbed mainly in the intrinsic region, thus increasing the collection probability of the generated carriers. Indeed, it can be seen in Figure 5 that for a fully etched patterned a-Si:H membrane, the electrical field intensity is the highest in the first tens of nanometres from the top surface, where the lifetime of the photo-generated carriers would be the shortest, and falls below 10% in the middle of the absorbing layer (the weak increase observed on the back side of the membrane is finally attributed to the reflection of the light on this rear surface). While it is important to drill sufficiently deep holes, care should be taken not to etch the silicon layer through the whole thickness. Indeed, due to the transmission phenomenon discussed above, a thin portion of the layer should be left unpatterned below the holes lattice, in order to absorb the remaining transmitted light. More precisely, by simply considering the Beer-Lambert law at 400 nm, it can be shown that keeping a 12 nm thin unpatterned a-Si:H layer is sufficient to absorb 50% of the light transmitted through the PhC structure, a fraction of light which would have been lost in the transparent and conductive oxide and in the highly doped layers for a planar layer covered by an anti-reflection coating. Considering the global light trapping strategy to optimize the efficiency of thin-film solar cells, it could therefore be possible to optimize simultaneously the light trapping effect through Fabry-Perot or guided mode resonances at large wavelengths, and the absorption at short wavelengths through the channelling effect discussed in this paper.

However, as the channelling regime reported here leads to a strong absorption in the first tens of nanometers of the sidewalls, it is expected to give rise to severe surface recombination. This should be circumvented by a conformal deposition of a passivation layer, possibly combined with the etching of the physically and chemically damaged regions, close to the sidewalls (see, e.g. [25]). Additionally, channelling could be the main effect responsible for the poor blue response in a-Si:H based radial junction solar cells as it is indeed not favourable if doped layers are located on the sidewalls, since the generated photocarriers then suffer from strong recombinations [26]. Finally, it appears that the selected doping scheme and passivation techniques should be taken into account while optimizing the nano-hole arrays geometrical parameters, in order to limit the negative impact of the surface recombination of the photocarriers [27], while taking advantage of the blue light channelling.

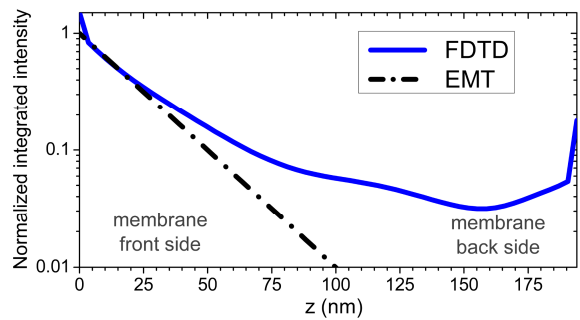

**Figure 5.** Electrical field intensity integrated over one PhC period and calculated with FDTD simulations or with the EMT approximation, as a function of the depth inside the 2D PhC patterned membrane. The EMT results were normalized to 1 by the field values at the top surface, while the FDTD data were normalized to 1 by the field value at the top surface but after removing all top surface effects, leading to values exceeding 1 on the front side of the membrane.

As a conclusion, we have demonstrated that the role of a PhC structure patterned in an absorbing layer is not limited to light harvesting at long wavelengths using optical resonances like guided Bloch modes. Indeed, it is also very efficient to couple incident light into vertically channelling modes at short wavelengths, increasing the absorption up to 45% relatively to the unpatterned sample. This remarkable channelling effect is characterized by an optical transmission across the investigated PhC structures. In the present case, it could be clearly demonstrated that more than 90% of the non-reflected light could be coupled to channelling modes below 500 nm, to be finally absorbed in the patterned a-Si:H layer or transmitted through the membrane. This optical regime is believed to impact the electrical performances of various configurations of nano-patterned thin-film solar cells [28, 29]. It can be leveraged to increase the external quantum efficiency of solar cells in the blue region of the useful spectrum by improving the absorption profile within the patterned active layer.

Funding Sources

This work was supported by Orange Labs Networks contract 0050012310-A09221 and the Repcop project of the LabEx iMUST of Université de Lyon.

ACKNOWLEDGMENT
The patterned samples used in this study were fabricated at Nanolyon technological facility, with the kind assistance of Pierre Cremillieu and Radoslaw Mazurczyk. The authors would like to thank Aziz Benamrouche from INL for his help in the topographical characterization of the samples. Serge Huant and Jean-François Motte from the Néel Institute are acknowledged for providing the SNOM fibres.